\documentclass[conference]{IEEEtran}

\PassOptionsToPackage{table}{xcolor}
\usepackage{url}
\usepackage{enumitem}
\usepackage{array}
\usepackage{booktabs}
\usepackage{xspace}
\usepackage{tikz}
\usepackage{tabularx}
\usepackage{xcolor}
\usepackage{microtype}
\usepackage[hidelinks,breaklinks=true]{hyperref}

\microtypesetup{expansion=false}
\usetikzlibrary{positioning,shapes.geometric,arrows.meta,calc,backgrounds,decorations.markings}

\definecolor{privRed}{HTML}{B2182B}
\definecolor{privBg}{HTML}{FDECEA}
\definecolor{pubBlue}{HTML}{4477AA}
\definecolor{pubBg}{HTML}{EEF3FA}
\definecolor{teal}{HTML}{228833}
\definecolor{tealBg}{HTML}{E6F4F1}
\definecolor{neutralGray}{HTML}{4A5568}
\definecolor{midgray}{HTML}{666666}
\definecolor{stagefill}{HTML}{F5F5F5}
\definecolor{darkgray}{HTML}{3D4349}
\definecolor{warnBorder}{HTML}{C0490A}
\definecolor{warnBg}{HTML}{FEF3EB}
\definecolor{warnText}{HTML}{7D3000}
\definecolor{passGreen}{HTML}{228833}
\definecolor{passBg}{HTML}{EBF7EF}
\definecolor{armOrange}{HTML}{9A3A00}
\definecolor{armOrangeBg}{HTML}{FEF0E6}
\definecolor{axisGray}{HTML}{667085}
\definecolor{linkGray}{HTML}{8A94A6}
\definecolor{dashgray}{HTML}{9BA4AF}
\definecolor{armZero}{HTML}{1A5276}
\definecolor{armZeroBg}{HTML}{E8F0FB}
\definecolor{armPolicy}{HTML}{117864}
\definecolor{armPolicyBg}{HTML}{E6F4F1}
\definecolor{humanGray}{HTML}{6F7782}
\definecolor{humanBg}{HTML}{F2F4F7}
\definecolor{execblue}{HTML}{4477AA}

\newcolumntype{L}{>{\raggedright\arraybackslash}X}
\newcolumntype{C}{>{\centering\arraybackslash}X}

\newcommand{\ValPilotLlmZeroViolationRate}{0.4500}
\newcommand{\ValPilotLlmPolicyViolationRate}{0.6700}

\newcommand{\ValPilotLlmZeroRunCount}{400}

\newcommand{\ValPilotLlmZeroHardViolationCount}{186}
\newcommand{\ValPilotLlmZeroViolationCiLow}{0.4020}
\newcommand{\ValPilotLlmZeroViolationCiHigh}{0.4990}
\newcommand{\ValPilotLlmZeroEnforcementModRate}{0.4500}

\newcommand{\ValPilotLlmPolicyRunCount}{400}

\newcommand{\ValPilotLlmPolicyHardViolationCount}{280}
\newcommand{\ValPilotLlmPolicyViolationCiLow}{0.6225}
\newcommand{\ValPilotLlmPolicyViolationCiHigh}{0.7143}
\newcommand{\ValPilotLlmPolicyEnforcementModRate}{0.6700}

\newcommand{\ValPilotHumanRunCount}{200}
\newcommand{\ValPilotHumanRunsWithViolations}{0}
\newcommand{\ValPilotHumanHardViolationCount}{0}

\newcommand{\ValPilotHumanViolationCiHigh}{0.0188}

\newcommand{\ValGateIncidentCount}{200}

\newcommand{\ValCampaignModelCount}{2}
\newcommand{\ValCampaignArmCount}{2}
\newcommand{\ValGateHumanBaselineCount}{200}
\newcommand{\ValGateLlmTrajectoryCount}{800}
\newcommand{\ValGateLlmTrajectorySuccessCount}{800}
\newcommand{\ValGatePreflightFailureCount}{0}
\newcommand{\ValGateExecutionFailureCount}{0}
\newcommand{\ValGateRunsPerRow}{200}

\newcommand{\ValGateCampaignProjectedLlmTrajectories}{800}

\newcommand{\ValOfficialClaudeZeroViolationRate}{0.3600}
\newcommand{\ValOfficialClaudePolicyViolationRate}{0.8700}
\newcommand{\ValOfficialGptZeroViolationRate}{0.5400}
\newcommand{\ValOfficialGptPolicyViolationRate}{0.4700}
\newcommand{\ValOfficialClaudeZeroDeltaJaccard}{0.0678}
\newcommand{\ValOfficialClaudeZeroPrecisionRaw}{0.8268}
\newcommand{\ValOfficialClaudeZeroPrecisionEnforced}{0.9454}
\newcommand{\ValOfficialClaudePolicyDeltaJaccard}{0.1544}
\newcommand{\ValOfficialClaudePolicyPrecisionRaw}{0.6673}
\newcommand{\ValOfficialClaudePolicyPrecisionEnforced}{0.9271}
\newcommand{\ValOfficialGptZeroDeltaJaccard}{0.1092}
\newcommand{\ValOfficialGptZeroPrecisionRaw}{0.8071}
\newcommand{\ValOfficialGptZeroPrecisionEnforced}{0.9792}
\newcommand{\ValOfficialGptPolicyDeltaJaccard}{0.0917}
\newcommand{\ValOfficialGptPolicyPrecisionRaw}{0.8108}
\newcommand{\ValOfficialGptPolicyPrecisionEnforced}{0.9733}
\newcommand{\ValOfficialEnforcementModificationRate}{0.5600}

\newcommand{\ValOfficialRemovedActionCount}{466}
\newcommand{\ValOfficialViolationCountRThree}{431}
\newcommand{\ValOfficialViolationCountRFour}{35}

\newcommand{\ValOfficialLlmOnlyRestoreHostCount}{431}
\newcommand{\ValOfficialLlmOnlyBlockEgressCount}{90}
\newcommand{\ValOfficialLlmOnlyIsolateHostCount}{35}
\newcommand{\ValOfficialLlmOnlyResetAdminCount}{14}
\newcommand{\ValOfficialHumanOnlyResetAdminCount}{515}
\newcommand{\ValOfficialHumanOnlyIsolateHostCount}{43}
\newcommand{\ValStabilityRepeatCount}{3}

\newcommand{\ValStabilityIncidentViolationRateMin}{0.8700}
\newcommand{\ValStabilityIncidentViolationRateMax}{0.8750}

\newcommand{\ValStabilityEnforcementModificationRateMin}{0.5487}
\newcommand{\ValStabilityEnforcementModificationRateMax}{0.5613}

\newcommand{\ValStabilityViolationCountRThreeMin}{422}
\newcommand{\ValStabilityViolationCountRThreeMax}{435}
\newcommand{\ValStabilityViolationCountRFourMin}{35}

\newcommand{\ValCorpusIncidentCount}{200}

\newcommand{\ValCorpusIncidentTypeCount}{10}
\newcommand{\ValCorpusIncidentTypeSummary}{access=7, authentication=4, exploit=19, falcon\_detection\_method=4, malware=76, policy=3, potential\_exploit=4, recon=2, suspicious\_activity=73, system=8}
\newcommand{\ValCatalogActionCount}{5}

\newcommand{\ValMappingActionCount}{3}
\newcommand{\ValMappingActionCoverageOverCatalog}{0.6000}

\newcommand{\ValCorpusMappingCoverage}{1.0000}
\newcommand{\ValCorpusMappedTaskCount}{1147}
\newcommand{\ValCorpusTaskCount}{1147}
\newcommand{\ValCorpusUnmatchedTaskCount}{0}

\newcommand{\ValCorpusSensitivityTaskCount}{1147}
\newcommand{\ValCorpusAmbiguousMatchCount}{0}

\newcommand{\ValCorpusSingleKeywordMatchCount}{579}

\newcommand{\ValBaselineUnsupportedActionIds}{block\_egress\_ip, restore\_host}
\newcommand{\ValApprovalProxyCoveredActionIds}{isolate\_host}
\newcommand{\ValApprovalProxyMissingActionIds}{restore\_host}
\newcommand{\ValSupportCollectForensicsIncidentCount}{200}
\newcommand{\ValSupportCollectForensicsMatchCount}{765}
\newcommand{\ValSupportCollectForensicsSingleKeywordShare}{0.7477}
\newcommand{\ValSupportIsolateHostIncidentCount}{191}
\newcommand{\ValSupportIsolateHostMatchCount}{191}

\newcommand{\ValSupportResetAdminIncidentCount}{191}
\newcommand{\ValSupportResetAdminMatchCount}{191}

\newif\ifartifactanonymous
\artifactanonymoustrue

\providecommand{\Description}[1]{}

\begin{document}

\title{SOCpilot: Verifying Policy Compliance for LLM-Assisted Incident Response}

\author{\IEEEauthorblockN{Sidnei Barbieri$^{*}$, Leonardo Vaz de Meneses$^{*}$, \\ Ágney Lopes Roth Ferraz$^{*}$ and Lourenço Alves Pereira Júnior$^{*}$}
\IEEEauthorblockA{$^{*}$Aeronautics Institute of Technology\\
Email: sbarbier@andrew.cmu.edu, leonardomeneses@ita.br, roth@ita.br, ljr@ita.br}

}

\maketitle

\begin{abstract}
Security operations centers (SOCs) are beginning to use large language models (LLMs) as copilots to draft incident-response plans. These plans may include actions that are valid per the catalog but still violate mandatory steps, required ordering, or approval gates before analyst review. SOCpilot makes this compliance question measurable at the plan boundary. It fixes the incident package, action catalog, policy rules, verifier, and public evidence surface, then verifies the copilot's proposed action trace. We evaluate two LLM providers on \ValGateIncidentCount{} real incidents from an anonymized production SOC in a financial-sector case study. We compare their plans to paired analyst-authored baselines from the same security orchestration, automation, and response (SOAR) cases. An identical inline policy text moves the two providers in opposite directions. A deterministic verifier removes \ValOfficialRemovedActionCount{} non-compliant, approval-gated actions without reducing baseline task recall. Aggregate rates remain stable across \ValStabilityRepeatCount{} reruns of the fixed corpus. The official evidence focuses on approval-gated decisions regarding recovery and containment. The artifact also includes public activation checks for mandatory and ordering repairs. We release the runnable artifact\footnote{\url{https://github.com/c2dc/socpilot-artifact}} so independent reviewers can rederive the public results without access to private incident data.
\end{abstract}

\IEEEpeerreviewmaketitle

\section{Introduction}
\label{sec:introduction}

Security operations centers (SOCs) are increasingly using large language models (LLMs) as copilots for incident response. In this paper, a copilot is a non-autonomous proposal component. It reads a canonical incident package and recommends a sequence of catalog actions for analyst review. The copilot does not execute any tools, approve any actions, or alter any hosts. This boundary is important because a plausible recommendation can still be policy-noncompliant. For example, restoring a host before preserving evidence can destroy forensic opportunity. Isolating a production asset without approval can interrupt business operations. Delaying containment can leave an attacker active~\cite{kokulu2019matched,alahmadi2022falsepositives,vermeer2023alertalchemy}.

The concern is not hypothetical. Field studies of SOCs report that legal, insurance, coordination, and organizational pressures constrain which response actions an organization can safely take~\cite{woods2023lessonslost}. Analysts often diverge from documented playbook steps in systematic ways~\cite{schlette2024playbooks}. Meanwhile, recent work has extended LLM assistance to security tasks such as penetration testing~\cite{deng2024pentestgpt}, goal-directed web scanning~\cite{stafeev2025yurascanner}, and cybersecurity knowledge assessment~\cite{wang2025digitalexpert}. The missing-measurement question is direct: when an LLM proposes an incident-response plan, does the action sequence satisfy the policy rules governing the analyst’s next decision?

Current evaluation practice makes it hard to answer that question. A reported ``LLM policy-compliance’’ result can vary depending on the prompt, incident slice, action vocabulary, policy interpretation, provider endpoint, or private post-processing. Runtime-control systems such as Progent~\cite{shi2025progent} and AgentSpec~\cite{wang2025agentspec} make constraints explicit for live agents by controlling tool calls. However, copilot deployments expose an earlier boundary: the analyst reads and approves the plan before any tool is invoked. The plan, not the tool call, is therefore the first place where a policy-noncompliant recommendation can do harm. SOCpilot targets that boundary. If the plan is the object the analyst sees, then the plan must be the object the evaluation verifies.

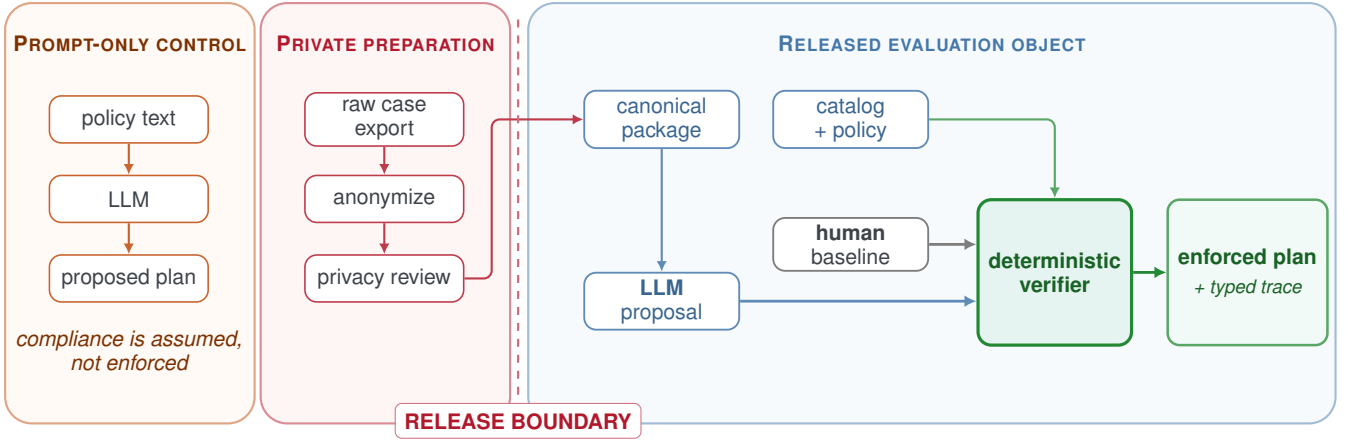
\begin{figure*}[htpb]
\centering
\makebox[\textwidth][c]{% SOCpilot's visual contract.
% Three panels: PROMPT-ONLY CONTROL (status quo), PRIVATE PREPARATION (not
% released), RELEASED EVALUATION OBJECT. The figure makes three
% relationships visible: (i) prompt-only control assumes compliance; (ii)
% private preparation stops at the release boundary; (iii) the released
% object composes canonical packages, the action catalog and policy, LLM
% proposals, paired human baselines, the deterministic verifier, and its
% enforced-plan trace.
\begin{tikzpicture}[
  font=\sffamily\footnotesize,
  >={Latex[length=4pt,width=4pt]},
  panel/.style={draw, rounded corners=10pt, line width=0.8pt, inner sep=0pt},
  ctrlbox/.style={
    draw=warnBorder!85, fill=white, text=darkgray,
    rounded corners=4.5pt, line width=0.6pt,
    minimum width=2.10cm, minimum height=0.62cm,
    align=center, inner sep=2.5pt, font=\sffamily\footnotesize
  },
  prepbox/.style={
    draw=privRed!85, fill=white, text=darkgray,
    rounded corners=4.5pt, line width=0.6pt,
    minimum width=2.10cm, minimum height=0.62cm,
    align=center, inner sep=2.5pt, font=\sffamily\footnotesize
  },
  pubbox/.style={
    draw=pubBlue!85, fill=white, text=pubBlue!85!black,
    rounded corners=4.5pt, line width=0.7pt,
    minimum width=2.05cm, minimum height=0.66cm,
    align=center, inner sep=2.5pt, font=\sffamily\footnotesize
  },
  refbox/.style={
    draw=midgray!85, fill=white, text=darkgray,
    rounded corners=4.5pt, line width=0.7pt,
    minimum width=2.05cm, minimum height=0.66cm,
    align=center, inner sep=2.5pt, font=\sffamily\footnotesize
  },
  verifierbox/.style={
    draw=teal, fill=tealBg, text=teal!72!black,
    rounded corners=5pt, line width=1.2pt,
    minimum width=1.82cm, minimum height=1.92cm,
    align=center, inner sep=4pt,
    font=\sffamily\bfseries\footnotesize
  },
  outbox/.style={
    draw=teal!75, fill=tealBg!50, text=teal!72!black,
    rounded corners=4.5pt, line width=0.95pt,
    minimum width=2.02cm, minimum height=1.92cm,
    align=center, inner sep=4pt, font=\sffamily\footnotesize
  },
  flow/.style={line width=0.8pt, -{Latex[length=4pt,width=4pt]}},
  thickflow/.style={line width=1.05pt, -{Latex[length=4.5pt,width=4.5pt]}},
  paramflow/.style={line width=0.8pt, -{Latex[length=4pt,width=4pt]}, draw=teal!75}
]

% =====================================================================
% PANEL BACKGROUNDS
% Total native width 17.55 cm. Layout keeps a visible release-boundary gap
% between private preparation and the released object.
% Heights: 5.55 cm to allow vertical breathing for internal boxes.
% =====================================================================
\begin{scope}[on background layer]
  \node[panel, draw=warnBorder!50, fill=warnBg!45,
        minimum width=3.30cm, minimum height=5.55cm, anchor=north west]
    (ctrlPanel) at (-8.80,2.65) {};
  \node[panel, draw=privRed!50, fill=privBg!50,
        minimum width=3.30cm, minimum height=5.55cm, anchor=north west]
    (prepPanel) at (-5.42,2.65) {};
  \node[panel, draw=pubBlue!50, fill=pubBg!45,
        minimum width=10.68cm, minimum height=5.55cm, anchor=north west]
    (pubPanel) at (-1.88,2.65) {};
\end{scope}

% Panel titles (uppercase smcaps, footnotesize for visual harmony with body).
\node[anchor=north, font=\sffamily\bfseries\footnotesize, text=warnText]
  at ([yshift=-9pt]ctrlPanel.north) {\textsc{Prompt-only control}};
\node[anchor=north, font=\sffamily\bfseries\footnotesize, text=privRed]
  at ([yshift=-9pt]prepPanel.north) {\textsc{Private preparation}};
\node[anchor=north, font=\sffamily\bfseries\footnotesize, text=pubBlue]
  at ([yshift=-9pt]pubPanel.north) {\textsc{Released evaluation object}};

% =====================================================================
% PROMPT-ONLY CONTROL nodes (vertical stack)
% Panel x range: -8.80 to -5.50, centre x = -7.15
% =====================================================================
\node[ctrlbox] (policyText) at (-7.15,1.10) {policy text};
\node[ctrlbox] (promptLlm)  at (-7.15,0.05) {LLM};
\node[ctrlbox] (promptPlan) at (-7.15,-1.00) {proposed plan};
\node[font=\sffamily\footnotesize\itshape, text=warnText, align=center]
  at (-7.15,-1.95) {compliance is assumed,\\not enforced};

\draw[flow, draw=warnBorder!85] (policyText.south) -- (promptLlm.north);
\draw[flow, draw=warnBorder!85] (promptLlm.south) -- (promptPlan.north);

% =====================================================================
% PRIVATE PREPARATION nodes (vertical stack)
% Panel x range: -5.42 to -2.12, centre x = -3.77
% =====================================================================
\node[prepbox] (rawExport)     at (-3.77,1.10)  {raw case\\export};
\node[prepbox] (anonymize)     at (-3.77,0.05)  {anonymize};
\node[prepbox] (privacyReview) at (-3.77,-1.00) {privacy review};

\draw[flow, draw=privRed!85] (rawExport.south) -- (anonymize.north);
\draw[flow, draw=privRed!85] (anonymize.south) -- (privacyReview.north);

% =====================================================================
% RELEASED EVALUATION OBJECT nodes
% Panel x range: -2.05 to 8.80 (width 10.85). Internal layout:
%   canonical/catalog top row; proposal/baseline lower row.
%   catalog+policy enters the verifier from above; proposals/baselines enter
%   from the left; verifier output flows right.
% =====================================================================
\node[pubbox] (canon)     at (-0.10,1.10)  {canonical\\package};
\node[pubbox] (catalog)   at ( 2.40,1.10)  {catalog\\+ policy};

\node[pubbox] (proposal)  at (-0.10,-1.30) {\textbf{LLM}\\proposal};
\node[refbox] (baseline)  at ( 2.40,-0.55) {\textbf{human}\\baseline};

\node[verifierbox] (verifier) at (5.10,-0.92)
  {deterministic\\verifier};
\node[outbox] (output) at (7.65,-0.92)
  {\textbf{enforced plan}\\[2pt]{\sffamily\scriptsize\itshape\mdseries + typed trace}};

% Inputs feed proposal sources.
\draw[flow, draw=pubBlue!85] (canon.south)   -- (proposal.north);

% Proposal sources feed the verifier from the left without crossing labels.
\draw[thickflow, draw=pubBlue!85]
  (proposal.east) -- (proposal.east -| verifier.west);
\draw[thickflow, draw=midgray!85]
  (baseline.east) -- (baseline.east -| verifier.west);

% Catalog/policy parameterizes the verifier from above.
\draw[paramflow, rounded corners=3pt]
  (catalog.east) -- ++(0.52,0)
  -| ($(verifier.north)+(0,0.40)$) -- (verifier.north);

% Verifier emits the enforced plan + typed trace (single output, no loop).
\draw[thickflow, draw=teal] (verifier.east) -- (output.west);

% =====================================================================
% Cross-boundary arrow: privacy review (private) -> canonical (public).
% =====================================================================
\draw[flow, draw=privRed!85, rounded corners=3pt]
  (privacyReview.east) -- ++(0.36,0) |- (canon.west);

% =====================================================================
% RELEASE BOUNDARY: a vertical boundary between private preparation and the
% released evaluation object. Prompt-only control is a comparison, not part of
% the release path.
% =====================================================================
\draw[draw=privRed!62, line width=0.85pt, dashed]
  (-2.00,2.42) -- (-2.00,-2.62);
\node[font=\sffamily\bfseries\footnotesize, text=privRed,
      fill=white, draw=privRed!60, rounded corners=2.5pt, inner sep=3.5pt]
  at (-2.00,-2.88) {RELEASE BOUNDARY};

\end{tikzpicture}}
\caption{\label{fig:pipeline_arch}SOCpilot’s declared evaluation object. Private data stops at the release boundary. The released object contains canonical packages, the action catalog and policy, LLM proposals, paired human baselines, and a deterministic verifier that emits enforced plans and typed traces. Together, these components define the surface on which every reported claim is checked.}
\Description{Three-panel architecture contrasting prompt-only control, private preparation, and the released evaluation object with canonical packages, catalog, and policy, LLM proposals, paired human baselines, deterministic verifier, and enforced plan traces.}
\end{figure*}

SOCpilot (Figure~\ref{fig:pipeline_arch}) turns this boundary into a fixed, auditable evaluation object. Each run uses one canonical incident package, one action catalog, one typed policy set, and one deterministic verifier path. With the same public inputs, a reviewer can see what the copilot proposed, which policy rule fired, what repair the verifier applied, and which aggregate claim follows. The private SOC data stay behind the release boundary. The public artifact contains the canonical packages, the mapping contract, the verifier, run accounting, and aggregate analyses needed to rederive the reported results.

SOCpilot is a sector-agnostic evaluation methodology. The reported experiment is a financial-sector case study. The artifact consumes canonical incident packages and a declared policy surface. Another SOC can replace the released cases with its own anonymized cases and rerun the same verifier. In the case study, external attackers trigger incidents against production assets. Responders operate under explicit constraints, institutional coordination, and a governed playbook~\cite{schlette2024playbooks,woods2023lessonslost,vermeer2023alertalchemy}. The unit of analysis is an incident-level response plan derived from a canonical SOAR case package. These are paired with analyst-authored workflow traces from the same case.

This framing leads to three research questions:

\noindent\hangindent=2.7em\hangafter=1 \textbf{RQ1:} How frequently do LLM-generated plans violate typed SOC policy rules on real incidents?

\noindent\hangindent=2.7em\hangafter=1 \textbf{RQ2:} How do these violations compare to paired human baselines from the same incidents?

\noindent\hangindent=2.7em\hangafter=1 \textbf{RQ3:} To what extent does deterministic verification reduce violations while preserving plan utility?

Our central claim is that LLM-assisted incident-response planning should verify policy compliance on the proposed plan rather than treat compliance as a property of the prompt. One declared evaluation on \ValCorpusIncidentCount{} real SOC incidents supports this claim.

The first observation is provider fragility. Under one prompt lineage and one shared incident package, two production LLM providers produce action-level violation rates that diverge. We report this gap with Cohen’s $h$, an effect-size measure for the difference between two proportions: the same inline policy text produces a large adverse shift for \texttt{claude-sonnet-4-6} ($h=1.12$) and only a small shift for \texttt{gpt-5.2} ($|h|\approx 0.14$)~\cite{cohen1988statistical}. The second observation is enforcement stability at the action boundary. The verifier operates only on the explicit action sequence and removes \ValOfficialRemovedActionCount{} non-compliant approval-gated actions while preserving paired task recall against analyst-authored baseline plans for the same incidents. The third observation is protocol stability: across \ValStabilityRepeatCount{} reruns of the fixed corpus, the violation rate stays within \ValStabilityIncidentViolationRateMin{}--\ValStabilityIncidentViolationRateMax{}, the enforcement-modification rate within \ValStabilityEnforcementModificationRateMin{}--\ValStabilityEnforcementModificationRateMax{}, and no run loses any task coverage relative to its raw proposal.

Because the official provider's evidence concentrates on approval-gated rules, the artifact separately exposes public activation checks for mandatory and ordering repairs. This separation makes scope explicit and provides a stable reference point for later comparisons: replace the prompt, catalog, mapping contract, rule file, provider, or verifier, and re-evaluate against a declared object rather than a moving target. Figure~\ref{fig:pipeline_arch} summarizes the resulting evaluation contract.

\section{Threat Model}
\label{sec:threat-model}

LLM assistance introduces a pre-execution risk: the model may recommend actions that conflict with policy, such as restoring a host before forensic analysis or suggesting approval-gated actions without authorization. Such violations can lead to evidence loss, delayed containment, or unsafe recovery~\cite{kokulu2019matched,alahmadi2022falsepositives,vermeer2023alertalchemy}. Studies of prompt injection, prompt leakage, prompt inversion, and cross-app LLM exploitation show that LLM-integrated deployments also surface adversarial manipulation paths at the serving and integration layers~\cite{greshake2023notwhat,wu2025promptleakage,qu2025promptinversion,dong2025philosophers,wu2025isolategpt}. Wu et al.\ (IsolateGPT), for example, demonstrate that execution isolation between LLM-backed apps is necessary to prevent cross-app data leakage and instruction manipulation~\cite{wu2025isolategpt}. Our focus is on pre-execution plan compliance rather than autonomous tool execution or serving-layer defense. That focus motivates a typed policy layer that constrains proposed actions before execution rather than relying on prompt formulation for safety.

The threat model is the plan-review boundary before execution: the assistant neither executes tools nor alters hosts nor directly accesses approval systems. The main risk is that a non-compliant plan may appear plausible during review while violating evidence-preservation or approval constraints. Accordingly, the explicit action sequence, rather than the hidden model state, is the security-relevant unit. We therefore measure whether governance-relevant steps are made explicit in the plan, including typed prerequisites required by operational practice~\cite{schlette2024playbooks,woods2023lessonslost}.

\section{Policy Model and Verifier}
\label{sec:policy-model}

To make policy compliance measurable, we explicitly model policy enforcement. The verifier is a deterministic, typed repair engine over action sequences that encodes action-level policy invariants for proposed plans. Given the incident context and a proposed plan, the verifier produces an enforced plan with structured violation traces. Throughout the paper, a \emph{policy set} denotes the complete declared rule collection for the reported evaluation, and a \emph{rule} refers to a single entry in that set.

Policy is represented by three rule families: \texttt{mandatory}, \texttt{prohibit\_before}, and \texttt{require\_approval}. In the current implementation, the rule scope is at the incident and action levels; role and multi-entity constraints are not included. The global policy surface is defined by the declared action catalog and typed rule set, whereas observed rule activation depends on how these artifacts intersect with the fixed corpus.

The reported rules encode SOC governance points where policy treatment is operationally consequential. \texttt{R1} captures mandatory containment for a high-confidence reverse-shell pattern, not generic command execution, to avoid treating ambiguous administrative commands as containment failures. \texttt{R2} encodes evidence preservation before recovery. \texttt{R3} and \texttt{R4} encode approval control for recovery and containment actions that can change system state, interrupt business operations, or destroy forensic opportunity. These choices reflect the incident-response and playbook literature’s emphasis on evidence preservation, organizational coordination, and governed recovery paths in real SOC work~\cite{kokulu2019matched,schlette2024playbooks,woods2023lessonslost,vermeer2023alertalchemy}. The official corpus exercises approval governance more strongly than mandatory or ordering repair, since \texttt{R1} and \texttt{R2} require telemetry and sequencing patterns absent from the released slice. This activation pattern is a corpus property; public activation checks exercise the same verifier semantics over a richer candidate rule surface.

Operationally, each rule asks one well-formed question about the proposed action sequence. A \texttt{mandatory} rule asks whether an action must be present when telemetry matches a declared scope. A \texttt{prohibit\_before} rule asks whether a prerequisite action appears before a state-changing action. A \texttt{require\_approval} rule asks whether an approval-gated action has explicit approval evidence in the public canonical package. The verifier answers these questions by minimal repair: it inserts missing mandatory actions, inserts missing prerequisites before constrained actions, and removes unapproved approval-gated actions in the official mode. The alternative approval treatment, \texttt{defer\_to\_human\_approval}, is reported as a sensitivity path. Each repair emits a typed violation record. The verifier deterministically transforms the proposed action list into the policy-treated sequence used for measurement.

Table~\ref{tab:policy_surface} surfaces the declared action vocabulary and the rule treatment attached to each action. The action catalog contains five actions spanning forensics, containment, credential reset, network blocking, and recovery. \texttt{R1} uses a narrow reverse-shell scope predicate (\texttt{event\_type\_contains=command\_execution}, \texttt{command\_contains=[bash -i,/dev/tcp/]}); the released official corpus contains no matching telemetry, so \texttt{R1} contributes declared verifier semantics but no provider-run violations.

\begin{table}[htpb]
\caption{\label{tab:policy_surface}Declared action-policy surface for the official evaluation.}
\centering
\footnotesize
\renewcommand{\arraystretch}{1.08}
\setlength{\tabcolsep}{1.5pt}
\rowcolors{2}{stagefill}{white}
\begin{tabularx}{\columnwidth}{@{}>{\raggedright\arraybackslash}m{0.23\columnwidth}>{\centering\arraybackslash}m{0.16\columnwidth}>{\centering\arraybackslash}m{0.16\columnwidth}>{\raggedright\arraybackslash}X@{}}
\toprule
Action & \shortstack[c]{Approval\\gate} & \shortstack[c]{Paired\\baseline} & Rule treatment \\
\midrule
Collect forensics & no & yes & R2 declared; inactive \\
Isolate host & yes & yes & R1 declared; R4 fires \\
Reset admin credentials & no & yes & No policy rule \\
Block egress address & no & no & No rule \\
Restore host & yes & no & R2 declared; R3 fires \\
\bottomrule
\end{tabularx}
\end{table}

The verifier expects a structured list of action identifiers. In LLM mode, the adapter requires JavaScript Object Notation (JSON) with the \texttt{recommended\_actions} field. Out-of-catalog actions are filtered before policy enforcement and logged in the artifact outputs. This contract constrains syntax at the verifier boundary, not semantic appropriateness. Catalog-valid but operationally mismatched actions remain part of the measured policy and utility burden.

Consider an input plan that contains only \texttt{restore\_host} under two applicable rules: \texttt{R2} requires forensic collection before restoration, and \texttt{R3} marks restoration as approval-gated. Pass~2 inserts \texttt{collect\_forensics} and records a \texttt{prohibit\_before} violation; Pass~3 applies the configured approval treatment to host restoration, records \texttt{approval\_required}, and, under the primary reported mode, removes the restoration action. The enforced plan retains only the forensics step, and the verifier emits two typed traces, \texttt{R2/order\_violation} (insertion of \texttt{collect\_forensics}) and \texttt{R3/approval\_required} (removal of \texttt{restore\_host}), at the same level of evidence recorded for every official run.

The procedure is finite, bounded, and deterministic under fixed inputs. Four implementation properties are checked by design: termination over finite action sequences, bounded repair work under the fixed rule set, idempotence after enforcement, and deterministic arbitration through a fixed pass order. Mandatory and approval-gate rules are applied once per pass, whereas ordering constraints are repaired to a fixed point after fail-fast checks reject self-dependencies and cycles. As a consistency check on the released policy, reapplying enforcement to already-enforced plans produces no further modifications within the active rule subset (\texttt{R3}, \texttt{R4}). This places the reported system in an intentionally narrow and auditable region of the design space: deterministic repair over typed action traces rather than global satisfiability solving over a richer workflow language.

\section{Verification Architecture}
\label{sec:system-architecture}

A system-level question remains beyond the policy framework: where to place the verifier relative to private incident handling and the public evaluation path. Figure~\ref{fig:pipeline_arch} depicts that flow. From left to right, raw exports and anonymization remain private, canonical packages cross the release boundary, and both LLM proposals and paired human baselines follow the same public verifier path. The result is a single enforcement layer applied to two proposal sources, with the release boundary positioned after anonymization rather than after model execution.

The verifier is implemented as a rule engine, not as another LLM. Three designs were considered. Prompt-only guidance keeps compliance in model and avoids a separate enforcement layer, but produces no repeatable audit trail and is sensitive to prompt drift across provider versions or model updates. An LLM judge introduces a structured verdict and a separate auditing model, but replaces deterministic enforcement with a second, non-deterministic decision layer, complicating cross-run comparisons and the audit of the verdict itself. Deterministic rule enforcement keeps the enforcement decision outside the model, produces a typed violation trace for every run, and is insensitive to provider drift under fixed policy artifacts. We chose the deterministic verifier for these repeatability and auditability properties, and for the operational simplicity it provides in a declared evaluation state. Solver-backed enforcement remains a relevant extension for richer policy semantics. This approach aligns with the monitor-based enforcement tradition in security policy research, where deterministic monitors serve as the final decision layer~\cite{schneider2000enforceable,ligatti2005editautomata,kim2015kinetic,yu2017psi}.

The system-boundary decision is central to the contribution. In our setting, the assistant does not execute tools directly, so the reviewed action sequence serves as the control point shared among providers, analysts, and the public artifact. Deterministic enforcement makes that boundary versionable and auditable even as provider behavior or prompt wording changes upstream.

The private/public split is determined by data governance requirements. In the case-study SOC, raw incident exports cannot leave the institution, in accordance with formal process and legal-governance constraints for incident handling~\cite{schlette2024playbooks,woods2023lessonslost}. The anonymizer is therefore the sole bridge between private data and publishable artifacts. Placing the boundary after anonymization ensures that the verification and evaluation pipeline remains public and artifact-evaluable, without exposing operational data.

\section{Corpus and Release Boundary}
\label{sec:dataset-protocol}

Raw exports and irreversible mappings remain local. Public artifacts are generated only after anonymization and privacy checks. The released artifacts are authored policy inputs, not verbatim copies of incident traces. Three release gates make the corpus usable for public review: privacy scanning, parseability checks, and deterministic task-to-action mapping.

The full corpus is audited for coverage diagnostics and candidate gaps. In the current \ValCorpusIncidentCount{}-incident audit, this process yielded no new action-catalog candidates and one corpus-wide approval gap (\texttt{require\_approval(isolate\_host)}), which was added to the active policy only after manual review. The public package contains canonical incident packages, active global artifacts, the protocol manifest, non-private analysis outputs, and the code, scripts, and runbooks needed to rerun integrity checks, dataset audits, and public assessment rechecks. It excludes raw exports, private-stage logs, and irreversible local mappings. Each per-incident public object contains telemetry (detection events, labels, and indicators), anonymized case metadata, deterministically mapped human actions, and quality traces that enable auditing of the conversion.

The corpus and canonicalization pipeline originates in a production SOC that ingests correlated Security Information and Event Management (SIEM) and Endpoint Detection and Response (EDR) alerts into a Security Orchestration, Automation, and Response (SOAR) case-management platform. This platform embeds playbook logic inside incident records, so exports contain both detection evidence and workflow traces. In practice, incident-specific artifacts provide telemetry, case metadata, and mapped human-action traces, while global artifacts provide the action vocabulary, policy semantics, and the mapping contract used to interpret those traces across the corpus.

Each raw export is a closed incident case authored by a human analyst within the SOAR platform: every selected case in the corpus reached \texttt{phase\_id = "Complete"} with at least one analyst-executed task on record. The export carries three workflow signals: \texttt{playbooks} and action menus (template references), \texttt{extracted\_tasks} (analyst-executed work units recorded by the platform), and note/telemetry fields (incident evidence). The paired human baseline is reconstructed deterministically: each \texttt{extracted\_task} is passed through the released task-to-action mapping contract, which projects it onto the \emph{same} canonical action catalog the LLM is constrained to use. The LLM arm and the human arm enter the verifier under one shared action vocabulary derived from the same incident package; they differ only in how the action sequence was authored (LLM proposal vs.\ recorded analyst execution). \texttt{playbooks} and platform action menus are not used to construct the baseline; they appear in the export only as contextual metadata. Mapping completeness is validated by parseability, coverage, and unmatched-task reporting, not by enforcing a minimum task count.

The fixed corpus consists of \ValCorpusIncidentCount{} real incidents selected from a production SOC export under three fixed inclusion criteria. The unit of analysis is a single closed incident case in the SOAR case management platform. Each selected case had to be closed, expose structured \texttt{extracted\_tasks} that support deterministic task-to-action mapping, and preserve cross-category diversity rather than collapsing into a single alert type. This is a purposive measurement corpus, not a prevalence sample of all SOC alerts: the design prioritizes per-incident measurement depth and cross-category diversity over prevalence estimation. The released slice spans \ValCorpusIncidentTypeCount{} incident categories (\ValCorpusIncidentTypeSummary) and exhibits operationally realistic severity variation across the low-through-critical range observed in the source SOC.

Each incident contributes an instrumented case package: telemetry, case metadata, mapped human actions, and the evidence needed to audit conversion quality. This package enables the public verification path without exposing raw operational data. The corpus should therefore be read as \ValCorpusIncidentCount{} instrumented evaluation cases from one anonymized institution, not as a claim about that institution’s full incident distribution or action-space coverage. Each case also incurs paired human-baseline construction, independent LLM plan generation across multiple model and prompt-arm combinations, deterministic policy verification, and multi-metric evaluation. This depth-over-breadth tradeoff is consistent with controlled experimental designs in security measurement~\cite{olszewski2025replicability}.

The abstraction is intentional. Playbook studies show that operational response procedures combine generic technical steps with organization-specific context, escalation paths, and local constraints, whereas incident-response field studies demonstrate that legal, coordination, and workflow pressures shape what can safely be done in a case~\cite{schlette2024playbooks,woods2023lessonslost,vermeer2023alertalchemy}. The canonical package preserves the evidence needed to evaluate one incident-level plan under a declared action contract, without inferring unshared context from guessed metadata. That discipline keeps the released corpus aligned with the public claim surface rather than silently treating the private workflow context as observable.

Raw incident exports are processed outside the public repository by a local private conversion stage that tokenizes sensitive fields, emits a canonical package, and records conversion metadata. Before any public artifact is released, the pipeline replaces email addresses, Internet Protocol (IP) addresses, phone numbers, hostnames, and user identifiers. Anonymization follows a dual-review protocol: the SOC operator performs initial sanitization and review in the operational environment, and the research team independently reprocesses the exported records through automated anonymization with personally identifiable information (PII) scanning, forbidden-term checks, and manual verification. This two-person, two-toolchain process reduces the risk of residual identifiers surviving either stage alone and makes the public release possible without exposing raw institution-internal incident data.

The primary risks are privacy leakage through logs, prompts, or snapshots; run contamination from stale artifacts; and non-deterministic preprocessing that could weaken traceability and public verification. Mitigations include the dual-review protocol described above, PII and forbidden-term scanning, strict dataset audit, release-boundary checks, and conversion manifests with quality metrics. The same release boundary also defines the public evidence surface for the paper: canonical incident packages for the active evaluation state, the declared action catalog and typed policy set, the task-to-action mapping contract used to derive paired human baselines, the manifest, shipped non-private analysis outputs, the public run-accounting manifest for the reported official evaluation, and the code and runbooks required to validate the released boundary. This declared surface is what artifact evaluators and future benchmark authors should treat as the checkable public object of the study.

\section{Evaluation Design}
\label{sec:method}

Given the fixed corpus and global artifacts described above, the evaluation unit consists of one incident, one model, one prompt arm, and one repeat; each incident also contributes one paired human baseline, evaluated using the same verifier and policy set. The active evaluation combines \ValGateIncidentCount{} incidents, \ValCampaignModelCount{} providers, \ValCampaignArmCount{} prompt arms, and one repeat, yielding \ValGateCampaignProjectedLlmTrajectories{} planned LLM trajectories and \ValGateHumanBaselineCount{} paired baselines. Table~\ref{tab:protocol_freeze_summary} consolidates the declared object end-to-end.

The design separates proposal variability from enforcement variability. Differences across models, arms, and reruns are part of the measured phenomenon and are reported as such, whereas enforcement variability is expected to collapse to zero under fixed inputs and policy artifacts. Rerun stability reflects the declared evaluation object rather than the universal determinism of providers. In the language of applied replicability, allowed variability belongs to the hypothesis, while fixed elements define the protocol~\cite{olszewski2025replicability}.

\begin{table}[htpb]
\caption{\label{tab:protocol_freeze_summary}Declared evaluation object for the official study.}
\centering
\footnotesize
\renewcommand{\arraystretch}{1.08}
\setlength{\tabcolsep}{3pt}
\rowcolors{2}{stagefill}{white}
\begin{tabularx}{\columnwidth}{@{}>{\raggedright\arraybackslash}p{0.30\columnwidth}>{\raggedright\arraybackslash}X@{}}
\toprule
Component & Frozen value \\
\midrule
Public corpus & \ValCorpusIncidentCount{} canonical incidents \\
Official slice & \ValGateIncidentCount{} incidents \\
Model set & \ValCampaignModelCount{} providers \\
Prompt arms & \ValCampaignArmCount{} arms \\
LLM trajectories & \ValGateCampaignProjectedLlmTrajectories{} planned; \ValGateLlmTrajectoryCount{} completed \\
Human baselines & \ValGateHumanBaselineCount{} paired \\
Stability reruns & \ValStabilityRepeatCount{} reruns \\
Active rule slice & approval gates (\texttt{R3}, \texttt{R4}) \\
\bottomrule
\end{tabularx}
\end{table}

Table~\ref{tab:protocol_freeze_summary} clarifies the measurement object, since the paper evaluates a declared evaluation state rather than a free-form assistant deployment: the incident slice, global artifacts, model set, and active comparison arms are all fixed before any outcome is counted.

In the current evaluation slice, the exercised rules are the approval-gated actions \texttt{R3} and \texttt{R4}, while \texttt{R1} and \texttt{R2} remain defined but inactive. Results should therefore be read as measurements over approval-governed behavior under the declared protocol, not over the full typed rule surface.

Two reporting rules follow from this freeze. First, top-line violation rates carry meaning only in the context of the actually exercised rule slice and its repair distribution. A low rate under approval-only activation is not equivalent to a low rate under a richer policy surface that also exercises \texttt{mandatory} insertion or \texttt{prohibit\_before} repair. Second, paired-baseline deltas remain comparable only as long as the mapping contract, action catalog, and verifier semantics are preserved. In measurement terms, these conditions define the hypothesis boundary, not ancillary implementation detail~\cite{olszewski2025replicability}.

Primary outcomes are \emph{run-level violation rate} (fraction of runs containing any policy violation), \emph{hard-violation burden} (hard violations per run, measuring within-run edit depth), and \emph{enforcement-modification rate} (fraction of runs where the verifier modifies the proposed plan, estimating safety-layer intervention burden). Secondary outcomes are \emph{task coverage} (recall against the paired human baseline after enforcement) and \emph{$\Delta$Jaccard} (change in action-set overlap after enforcement; positive values indicate movement toward the paired human baseline).

\subsection{Provider Execution}

For the active evaluation, LLM generation is minimally constrained. Policy enforcement is applied post-generation so that measurements capture raw policy-compliance risk in proposed plans, rather than the effect of upstream filters. Both arms share the same copilot contract: the model is a non-autonomous proposal component that returns a bounded action trace for analyst review, not an executable procedure or an approval authority. The user message contains incident metadata, a bounded telemetry sample, the action catalog with approval and reversibility metadata, an operational-rules block, and the required \texttt{recommended\_actions} JSON schema. The zero-arm leaves the operational-rules block empty; the policy-arm fills it with one line per active rule, including \texttt{rule\_id}, family, target action, severity, a prerequisite when applicable, repair operator, and a concise policy rationale. No provider receives a different policy surface. The only provider-specific formatting occurs at the application programming interface (API) level: OpenAI-compatible endpoints receive the system instruction as a chat message, whereas Anthropic receives the identical system text in the provider’s \texttt{system} field.

Recommended actions must resolve cleanly to catalog action identifiers via the JSON output contract. Suggestions that do not resolve, such as synonyms, composite labels, or obfuscated names, are logged as out-of-catalog and excluded from enforcement inputs rather than being coerced into uncertain semantic matches. In the official execution lineage, per-run prompt messages and prompt hashes are stored. The public artifact exposes the arm logic, model registry, mapping contract, arm-to-prompt mapping, and shared prompt template needed to audit the setup without rerunning provider calls.

The official cross-provider evaluation runs on \ValGateIncidentCount{} incidents under identical settings. The evaluation-facing model pair is \texttt{gpt-5.2} and \texttt{claude-sonnet-4-6}. All official runs use deterministic decoding (\texttt{temperature}=0.0, \texttt{max\_tokens}=512). Provider-declared model identifiers are recorded verbatim in run manifests and tables, so the paper reports the exact API labels used at execution time rather than retrospectively normalized names. The provider application programming interfaces do not expose fixed-seed control, so cross-run comparability is anchored by the declared evaluation state, manifests, and full run capture rather than by seed locking. To bound remaining serving-layer variance, the full fixed \ValGateIncidentCount{}-incident corpus was rerun \ValStabilityRepeatCount{} times under the same protocol, and the resulting stability range is reported in Section~\ref{sec:results}.

Hard violations are rule violations labeled \texttt{hard} in the policy set. In this study, these correspond to operationally unsafe recommendations such as missing mandatory containment steps, violating ordering constraints, or proposing approval-gated actions without approval context. The hard-violation rate is the fraction of runs that have at least one hard violation under the proposed plan. Because all active rules are labeled \texttt{hard}, the binary hard-violation rate coincides with the binary run-level violation rate, and evaluation tables also report \emph{hard violations per run} as the count-normalized burden metric.

\subsection{Paired Baseline and Endpoints}

Human incident actions define the paired human baseline for comparing LLM outputs on the same incidents. In the private stage, baseline actions are derived from SOAR \texttt{extracted\_tasks} via a deterministic task-to-action mapper with hash-tracked rules, producing ordered action identifiers in the public canonical package. This yields a deterministic workflow-level reference for overlap, burden, and compliance comparisons without claiming a full host-level execution trace. Thus, the paired human baseline is a mapped workflow reference under the same action contract, not a ground-truth execution trace. The action catalog, typed policy set, and task-to-action mapping rules remain authored inputs before execution; after freeze, canonicalization checks, LLM plan generation, deterministic verification, metric computation, and artifact rendering are automated.

Two truth anchors are used: the paired human baseline under the fixed-mapping contract and the verifier's decisions under the reported policy rules and action catalog. The primary endpoint is compliance under the declared policy, not host-level outcome truth or proof of policy optimality. The primary enforcement mode for approval-gated actions is \texttt{remove}. Incident-scoped approval evidence can be used when available, but in this corpus, approval annotations for mapped human actions are drawn from the task-to-action contract rather than from host-level approval logs. Therefore, the endpoint reflects compliance in the proxy-approval context, and \texttt{defer\_to\_human\_approval} remains available as an explicit sensitivity path, not the reported primary endpoint.

\subsection{Metrics and Statistics}

Violation outcomes are classified by policy-rule type and severity. Paired deltas in run-level violation rate and hard-violation burden (hard violations per run) are computed between LLM outputs and paired human baselines per incident. The verifier effect is measured as the reduction in hard-rule risk versus action preservation. Statistical reporting uses Wilson confidence intervals~\cite{wilson1927probable}, paired proportion tests, McNemar for matched incidents~\cite{mcnemar1947sampling}, Holm correction for multiple comparisons~\cite{holm1979sequentially}, and Cohen’s $h$ as a descriptive magnitude summary for marginal rate separation~\cite{cohen1988statistical}, following replicability-oriented reporting practices in modern security experimentation~\cite{olszewski2025replicability}. Beyond the binary run-level endpoint, per-run violation counts, severity counts, violation types, edit operations (insert, remove, reorder, defer), and action-overlap deltas versus paired human baselines are recorded, ensuring aggregation does not hide within-run edit burden. Inferential statistics are limited to contrasts among observed model/arm cells in the declared evaluation state, not to population-level claims across SOCs or policy spaces. The declared manifest fixes the policy rules, mapping rules, prompt templates, model identifiers, and primary outcomes; any changes must be reported as separate evaluation objects. To ensure traceability, mapping-rule Secure Hash Algorithm 256-bit (SHA-256) values are recorded in source manifests, prompt/input hashes are stored, and the paper’s tables and figures are rendered from the maintained analysis bundle and aggregate summaries shipped in the public artifact. Execution begins only after privacy, parseability, mapping quality, and manifest checks pass.

The active \ValGateIncidentCount{}-incident configuration contains one LLM trajectory for each incident, provider, and prompt arm, plus one paired human baseline per incident. This design gives \ValGateCampaignProjectedLlmTrajectories{} planned LLM trajectories and \ValGateHumanBaselineCount{} paired baselines. The realized evaluation completed \ValGateLlmTrajectoryCount{} LLM trajectories, with \ValGateExecutionFailureCount{} execution failures. Run accounting is preserved in the manifests, but it does not carry the compliance claim.

\section{Results}
\label{sec:results}

The released corpus of \ValCorpusIncidentCount{} real incidents supports end-to-end public audit and reruns at the released boundary. The main finding of the official evaluation is provider-specific within the approval-governed slice: under identical protocol conditions, prompt-level policy text improves compliance for \texttt{gpt-5.2} and degrades it for \texttt{claude-sonnet-4-6}, while the resulting burden remains concentrated in approval-gated actions within the approved catalog. The evidence is therefore strongest for the approval-governed slice, where the observed verifier action is removal, not insertion, reordering, or deferral. We begin with corpus release checks, then localize the main findings across provider-level rates, rule concentration, utility, and stability.

\subsection{Corpus Release Checks}

Corpus release quality is defined by the released conversion gates rather than by workflow completeness in the private environment. Across all \ValCorpusIncidentCount{} incidents, privacy and parseability checks passed, task extraction remained valid, and task-to-action mapping achieved complete released-surface coverage (\ValCorpusMappedTaskCount/\ValCorpusTaskCount{}) with zero unmatched tasks, zero ambiguous matches, and zero audited privacy issues. Under the declared protocol, these gates establish that the public corpus is fit for paired evaluation and reproducible reruns: the measurement object is well-formed, auditable, and free from conversion-induced distortion at the released boundary.

The same release evidence also bounds the claim. Mapper coverage over the full catalog is \ValMappingActionCoverageOverCatalog{} (\ValMappingActionCount/\ValCatalogActionCount{} catalog actions), reflecting human-task distribution rather than full action-space coverage. Paired-baseline support concentrates in three actions: \texttt{collect\_forensics} carries the deepest support (\ValSupportCollectForensicsIncidentCount{} cases, \ValSupportCollectForensicsMatchCount{} matches, of which \ValSupportCollectForensicsSingleKeywordShare{} of unique matches are single-keyword); \texttt{isolate\_host} (\ValSupportIsolateHostIncidentCount/\ValSupportIsolateHostMatchCount) and \texttt{reset\_admin\_credentials} (\ValSupportResetAdminIncidentCount/\ValSupportResetAdminMatchCount) carry stronger multi-keyword support; and \texttt{\ValBaselineUnsupportedActionIds} have no paired support in the reported slice. Approval-proxy support is similarly narrow: \texttt{\ValApprovalProxyCoveredActionIds} is covered in the mapping contract, whereas \texttt{\ValApprovalProxyMissingActionIds} remains approval-gated without paired baseline support. The released artifact therefore supports a specific claim about auditable policy enforcement over a fixed action interface, not a claim that the released slice exhausts the institution’s operational action space.

The pooled screening view is reported only as a protocol-level check, since the two providers move in opposite directions under policy prompting. Across both direct-provider models, the pooled arm rates are \ValPilotLlmZeroViolationRate{} for \texttt{llm\_zero} and \ValPilotLlmPolicyViolationRate{} for \texttt{llm\_policy\_prompt}, while the paired human baseline has no violating runs (\ValPilotHumanRunsWithViolations/\ValPilotHumanRunCount{}; 95\% confidence interval [0, \ValPilotHumanViolationCiHigh]). These pooled counts provide context but do not constitute a separate substantive claim, as the provider-level decomposition below drives the main interpretation. Two aggregate signals remain stable: violations are concentrated in approval-gated rules, and the stored plans contain no out-of-catalog actions under the reported output contract. This rules out syntactic catalog escapes in the stored plans, but not catalog-valid actions that may still misalign with the paired baseline.

\subsection{Provider-Level Compliance (RQ1)}

We then report the official \ValGateIncidentCount{}-incident evaluation. Under the fixed policy surface, the policy arm produces a borderline-detectable improvement for \texttt{gpt-5.2} (Cohen’s $|h|\approx 0.14$) and a large degradation for \texttt{claude-sonnet-4-6} ($h=1.12$). Across providers, all observed official violations fall in approval-gated rules, while deterministic enforcement reduces risk without reducing task coverage. The official run attempted \ValGateCampaignProjectedLlmTrajectories{} LLM trajectories and completed \ValGateLlmTrajectorySuccessCount{} (\ValGatePreflightFailureCount{} preflight failures; \ValGateExecutionFailureCount{} execution failures).

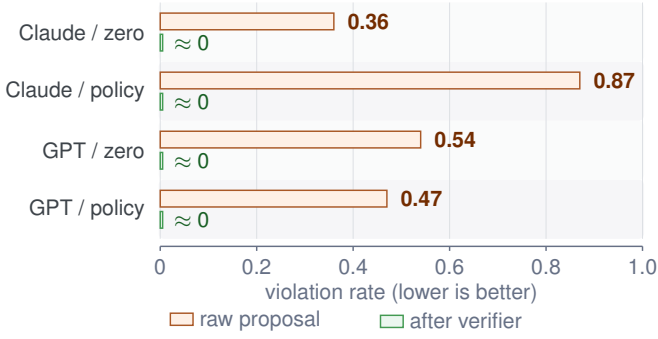
\begin{figure}[htpb]
\centering
% Single-column violation-rate chart: raw vs enforced bars per model/arm cell.
\begin{tikzpicture}[
  font=\sffamily\footnotesize,
  rawbar/.style={fill=armOrangeBg, draw=armOrange!85, line width=0.55pt},
  enfbar/.style={fill=passBg, draw=passGreen!85, line width=0.55pt},
  axisline/.style={draw=axisGray, line width=0.55pt},
  gridline/.style={draw=axisGray!22, line width=0.35pt}
]

% Geometry
\def\xscale{6.38}      % 1.0 rate unit -> 6.38 cm
\def\barH{0.22}
\def\halfBar{0.11}

% Row centres (top to bottom): claude/zero, claude/policy, gpt/zero, gpt/policy
% Each cell shows two bars vertically separated
\def\rowSep{0.28}     % vertical separation between raw and enforced bar centres

% Background alternating row shading (cell height 0.78)
\foreach \y/\sh in {0.30/6, 1.08/3, 1.86/6, 2.64/3}{
  \fill[axisGray!\sh, rounded corners=2pt]
    (-0.02,\y) rectangle (\xscale+0.02,\y+0.78);
}

% Vertical grid + x-axis
\foreach \x in {0,0.2,0.4,0.6,0.8,1.0}{
  \draw[gridline] (\x*\xscale,0.30) -- (\x*\xscale,3.42);
}
\draw[axisline] (0,0.20) -- (\xscale,0.20);
\foreach \x in {0,0.2,0.4,0.6,0.8,1.0}{
  \draw[axisline] (\x*\xscale,0.14) -- (\x*\xscale,0.20);
  \node[below, text=axisGray, font=\sffamily\footnotesize] at (\x*\xscale,0.16) {\x};
}
\node[text=axisGray, font=\sffamily\footnotesize] at (\xscale/2,-0.42)
  {violation rate (lower is better)};

% --- helper macro: draw one cell with raw+enforced pair ---
% #1 cell-centre y, #2 raw rate, #3 label
\newcommand{\cellrow}[3]{%
  % left label
  \node[anchor=east, text=darkgray] at (-0.08,#1) {#3};
  % raw bar above centre
  \fill[rawbar] (0,#1+\rowSep/2-\halfBar)
                rectangle (#2*\xscale,#1+\rowSep/2+\halfBar);
  \node[anchor=west, text=armOrange!75!black, font=\sffamily\bfseries\footnotesize]
    at (#2*\xscale+0.06,#1+\rowSep/2) {#2};
  % enforced bar below centre (rate ~0)
  \fill[enfbar] (0,#1-\rowSep/2-\halfBar)
                rectangle (0.005*\xscale,#1-\rowSep/2+\halfBar);
  \node[anchor=west, text=passGreen!70!black, font=\sffamily\footnotesize]
    at (0.06,#1-\rowSep/2) {$\approx$ 0};
}
\cellrow{3.03}{0.36}{Claude / zero}
\cellrow{2.25}{0.87}{Claude / policy}
\cellrow{1.47}{0.54}{GPT / zero}
\cellrow{0.69}{0.47}{GPT / policy}

% Inline two-entry legend just under x-axis label.
\node[anchor=west, text=axisGray, font=\sffamily\footnotesize] at (0,-0.78)
  {\tikz{\fill[rawbar] (0,0) rectangle (0.30,0.16);}~raw proposal};
\node[anchor=west, text=axisGray, font=\sffamily\footnotesize] at (2.78,-0.78)
  {\tikz{\fill[enfbar] (0,0) rectangle (0.30,0.16);}~after verifier};

\path (0,-0.96);

\end{tikzpicture}
\caption{\label{fig:gate2_violation_rates}Provider-level violation rates before and after deterministic verification. Raw proposals diverge under identical policy text; post-verifier rates collapse toward zero because the official violations are approval-gated.}
\Description{A horizontal bar chart with four model-arm rows: Claude/zero, Claude/policy, GPT/zero, and GPT/policy. Each row shows an orange raw-proposal bar and a thin green post-verifier bar that is essentially flat at zero.}
\end{figure}

The rule semantics are stated once in Table~\ref{tab:policy_surface}. The official evaluation exercises only the approval-centered slice: \texttt{R1} and \texttt{R2} remain inactive, and observed enforcement is removal rather than insertion, reordering, or deferral. Figure~\ref{fig:gate2_violation_rates} shows provider-level run-level violation rates (lower is better). The provider-level effect of policy prompting is asymmetric rather than a small fluctuation around a common trend: the same inline policy text increases violation prevalence for \texttt{claude-sonnet-4-6} while decreasing it for \texttt{gpt-5.2}.

Table~\ref{tab:evaluation_primary_outcomes} and Table~\ref{tab:evaluation_pairwise_tests} report per-cell rates and paired contrasts (\(n=\ValGateRunsPerRow\) per model/arm cell). All six paired contrasts survive Holm correction, but effect sizes vary widely. The three contrasts involving \texttt{claude-sonnet-4-6}/policy and the cross-provider zero-arm contrast (\texttt{claude}|zero vs.\ \texttt{gpt-5.2}|zero) carry the largest separations. The remaining two contrasts (\texttt{claude}|zero vs.\ \texttt{gpt-5.2}|policy and the within-\texttt{gpt-5.2} arm contrast) survive correction with effect sizes near \(|h|\!\approx\!0.14\), a magnitude small enough that the within-GPT effect should be read as borderline-detectable rather than substantively important. The figures and tables below quantify these claims.

\subsection{Rule Concentration and Enforcement (RQ2, RQ3)}

Together, these results separate proposal-level variability from enforcement-level stability. Figure~\ref{fig:gate2_violation_rates} captures the provider split in violation prevalence, while the outcome tables show that deterministic enforcement improves precision in every model/arm cell and preserves task coverage. Rule control follows directly from Table~\ref{tab:policy_surface}: in the official provider run, approval rules R3 and R4 account for all observed violations, while the \texttt{mandatory} and \texttt{prohibit\_before} rules never activate.

The concentration is specified by the protocol rather than being incidental. In the reported slice, approval rules bind to catalog actions proposed by both providers, whereas \texttt{R1} and \texttt{R2} require conditions absent from the released cases. The observed burden results from this intersection: the incident mix activates approval-gated actions in the approved catalog, while \texttt{mandatory} and \texttt{prohibit\_before} never receive the telemetry or sequencing evidence needed to fire.

\begin{table*}[htpb]
\caption{Official evaluation outcomes by model and arm. The shaded row is the paired human baseline; bold marks the highest violation rate.}
\label{tab:evaluation_primary_outcomes}
\centering
\footnotesize
\renewcommand{\arraystretch}{1.08}
\setlength{\tabcolsep}{3pt}
\rowcolors{2}{stagefill}{white}
\begin{tabularx}{\textwidth}{@{}>{\raggedright\arraybackslash}X
>{\centering\arraybackslash}p{0.125\textwidth}
>{\centering\arraybackslash}p{0.125\textwidth}
>{\centering\arraybackslash}p{0.125\textwidth}
>{\centering\arraybackslash}p{0.125\textwidth}
>{\centering\arraybackslash}p{0.125\textwidth}
>{\centering\arraybackslash}p{0.09\textwidth}
>{\centering\arraybackslash}p{0.09\textwidth}@{}}
\toprule
Provider / arm & Runs & \shortstack[c]{Violation\\rate} & \shortstack[c]{95\% Wilson\\interval} & \shortstack[c]{Hard\\violations/run} & \shortstack[c]{Verifier\\edit rate} & \shortstack[c]{Task\\coverage} & $\Delta$Jaccard \\
\midrule
Claude / zero & 200 & 0.3600 & [0.2967, 0.4286] & 0.3650 & 0.3600 & 0.7567 & 0.0678 \\
Claude / policy & 200 & \textbf{0.8700} & [0.8163, 0.9097] & 0.9100 & 0.8700 & 0.7634 & 0.1544 \\
GPT / zero & 200 & 0.5400 & [0.4708, 0.6077] & 0.5650 & 0.5400 & 0.8100 & 0.1092 \\
GPT / policy & 200 & 0.4700 & [0.4020, 0.5391] & 0.4900 & 0.4700 & 0.7400 & 0.0917 \\
\rowcolor{gray!8}
Human baseline & 200 & 0 & [0, 0.0188] & 0 & 0 & 1.0000 & 0 \\
\bottomrule
\end{tabularx}
\end{table*}

\begin{table*}[htpb]
\caption{Official paired contrasts by model and arm. Holm-adjusted McNemar tests and Cohen's $h$ summarize matched rate separation.}
\label{tab:evaluation_pairwise_tests}
\centering
\footnotesize
\renewcommand{\arraystretch}{1.08}
\setlength{\tabcolsep}{3pt}
\rowcolors{2}{stagefill}{white}
\begin{tabularx}{\textwidth}{@{}>{\raggedright\arraybackslash}X
>{\centering\arraybackslash}p{0.15\textwidth}
>{\centering\arraybackslash}p{0.15\textwidth}
>{\centering\arraybackslash}p{0.15\textwidth}
>{\centering\arraybackslash}p{0.15\textwidth}
>{\centering\arraybackslash}p{0.15\textwidth}@{}}
\toprule
Contrast & $n$ & $\Delta$ rate & McNemar $p$ & Holm $p$ & Cohen's $h$ \\
\midrule
Claude policy vs Claude zero & 200 & $+$0.5100 & \textless{}0.0001 & \textless{}0.0001 & $+$1.1169 \\
Claude policy vs GPT policy & 200 & $+$0.4000 & \textless{}0.0001 & \textless{}0.0001 & $+$0.8931 \\
Claude policy vs GPT zero & 200 & $+$0.3300 & \textless{}0.0001 & \textless{}0.0001 & $+$0.7530 \\
Claude zero vs GPT policy & 200 & $-$0.1100 & 0.0003 & 0.0006 & $-$0.2238 \\
Claude zero vs GPT zero & 200 & $-$0.1800 & \textless{}0.0001 & \textless{}0.0001 & $-$0.3639 \\
GPT policy vs GPT zero & 200 & $-$0.0700 & 0.0056 & 0.0056 & $-$0.1401 \\
\bottomrule
\end{tabularx}
\end{table*}

The tables anchor these visual patterns to exact per-cell rates and paired contrasts. Execution diagnostics reveal the operational decomposition of the repair burden: most removals are \texttt{restore\_host} approvals (R3), with a smaller, stable \texttt{isolate\_host} approval component (R4). Across the four provider/arm cells, all \ValGateCampaignProjectedLlmTrajectories{} planned LLM trajectories completed, and the stored official plans contain no out-of-catalog action identifiers.

Table~\ref{tab:utility_precision} decomposes overlap into raw and enforced precision, showing that deterministic enforcement consistently improves precision, while Table~\ref{tab:evaluation_primary_outcomes} confirms task coverage holds. The remaining utility gap is directional rather than diffuse. LLM-only actions concentrate in \texttt{restore\_host} (\ValOfficialLlmOnlyRestoreHostCount{}) and \texttt{block\_egress\_ip} (\ValOfficialLlmOnlyBlockEgressCount{}), with smaller contributions from \texttt{isolate\_host} (\ValOfficialLlmOnlyIsolateHostCount{}) and \texttt{reset\_admin\_credentials} (\ValOfficialLlmOnlyResetAdminCount{}); baseline-only actions concentrate in \texttt{reset\_admin\_credentials} (\ValOfficialHumanOnlyResetAdminCount{}) and \texttt{isolate\_host} (\ValOfficialHumanOnlyIsolateHostCount{}).

The action-level anatomy explains why the approval slice is operationally important despite its narrow rule coverage. The repair burden is driven primarily by recovery: \texttt{R3} accounts for \ValOfficialViolationCountRThree{} of the \ValOfficialRemovedActionCount{} removed approval-gated actions, whereas \texttt{R4} accounts for \ValOfficialViolationCountRFour{}. Enforcement improves overlap mainly by removing excess approval-gated recovery proposals, with containment contributing a smaller but still policy-relevant share. The \texttt{block\_egress\_ip} disagreement affects utility but not policy violation under the current rule set, since the action is catalog-valid and not approval-gated in this freeze.

\subsection{Rerun Stability}

Three observations summarize the official evaluation. First, the private-to-canonical conversion passed the parseability and privacy gates for all \ValCorpusIncidentCount{} incidents, and the task-to-action mapping achieved \ValCorpusMappedTaskCount/\ValCorpusTaskCount\ with weighted coverage \ValCorpusMappingCoverage. Across the official evaluation, run-level violation rates were \ValOfficialClaudeZeroViolationRate{} and \ValOfficialClaudePolicyViolationRate{} for \texttt{claude-sonnet-4-6} (zero vs.\ policy) and \ValOfficialGptZeroViolationRate{} and \ValOfficialGptPolicyViolationRate{} for \texttt{gpt-5.2}. In the declared evaluation state, prompt-level policy text does not uniformly improve compliance across providers.

Second, observed violations were concentrated entirely in approval-gated rules: \texttt{R3} produced \ValOfficialViolationCountRThree{} violations and \texttt{R4} produced \ValOfficialViolationCountRFour{}, with no out-of-catalog actions under the reported output contract. Deterministic enforcement modified \ValOfficialEnforcementModificationRate{} of LLM runs and removed \ValOfficialRemovedActionCount{} approval-gated actions; no run lost task coverage relative to its raw proposal, and the average \(\Delta\)Jaccard after enforcement remained positive in all four model/arm cells. The dominant observed failure mode is non-compliant action selection within the approved catalog, not catalog escape.

Third, action disagreement with the paired human baseline was directional: LLM-only actions were dominated by \texttt{restore\_host} (\ValOfficialLlmOnlyRestoreHostCount{}) and \texttt{block\_egress\_ip} (\ValOfficialLlmOnlyBlockEgressCount{}), while baseline-only actions were dominated by \texttt{reset\_admin\_credentials} (\ValOfficialHumanOnlyResetAdminCount{}). Across \ValStabilityRepeatCount{} completed repeated summaries of the fixed \ValGateIncidentCount{}-incident corpus, the aggregate picture remained stable: the incident-violation rate stayed within \ValStabilityIncidentViolationRateMin{}--\ValStabilityIncidentViolationRateMax{}, the enforcement-modification rate within \ValStabilityEnforcementModificationRateMin{}--\ValStabilityEnforcementModificationRateMax{}, no run lost task coverage in any rerun, \texttt{R3} stayed within \ValStabilityViolationCountRThreeMin{}--\ValStabilityViolationCountRThreeMax{} violations, and \texttt{R4} held fixed at \ValStabilityViolationCountRFourMin{}. Across providers and reruns, the same pattern holds: provider-level policy prompting and action-level policy adherence diverge under the declared protocol, and deterministic verification remains the last stable pre-execution control point over governed actions in the reported evaluation.

The discussion below interprets this as a systems result: SOCpilot makes plan-level policy compliance measurable at the action boundary and shows that deterministic enforcement is the stable control surface under the reported provider variation.

\section{Discussion}
\label{sec:discussion}

SOCpilot changes the deployment question from ``Did the prompt mention the policy?'' to ``Does the reviewed action trace satisfy the policy?'' The evaluation shows why this change matters. Under one declared prompt lineage and one shared incident package, prompt-level policy text did not yield stable action-level compliance across providers, whereas deterministic verification of typed action sequences did. The implication for system design is that the reviewed action trace, not the prompt, is the enforceable boundary.

The official provider evidence is concentrated in approval-governed actions (RQ1, RQ3). In the fixed corpus, \texttt{R1} depends on a high-confidence reverse-shell signature absent from the released incident slice, while \texttt{R2} requires restoration-before-forensics patterns not observed in the paired baselines or official LLM trajectories. This coverage result is part of the measurement: the paper reports which rule families generated provider evidence, and the artifact separately exercises richer repair modes outside the official estimate. The released artifact matters because it fixes what may vary across reruns and what must remain invariant when later studies claim improvement.

The main lesson from the data is that non-compliant action selection happens within the approved catalog. Out-of-catalog actions remain at zero under the reported output contract, yet provider behavior still diverges under inline policy text. The opposite prompting effects are informative: policy text can shift proposal distributions, but it does not guarantee action-level compliance. A degraded provider cell under inline policy text demonstrates that prompt-level policy exposure and action-level policy compliance are distinct. Conversely, zero out-of-catalog actions establishes syntax control, not operational adequacy. The measured burden remains inside the approved catalog, where governance-sensitive choices, such as restoration and containment, still require typed control before execution.

The protocol supports a concrete systems claim. Prompt-level policy text can shift proposal distributions, but the reviewed action trace is the enforceable boundary. The verifier is not an accessory to generation or a cleanup stage; it is the pre-execution control point once proposal variability reaches the action trace.

Operationally, approval concentration is central. Incident-response playbooks and studies of real SOC practice emphasize that containment and recovery steps are shaped by organizational coordination, approval paths, and external constraints, not just by detection evidence~\cite{schlette2024playbooks,woods2023lessonslost,vermeer2023alertalchemy}. Actions such as restoring a host or isolating it prematurely can alter the system state, disrupt business workflows, or compromise forensic evidence. The activated slice sits where governance pressure is highest.

Within the evaluated protocol, SOCpilot is a bounded proposal engine plus a deterministic policy layer. The copilot proposes catalog actions; the verifier enforces typed approvals and ordering constraints before analyst review; the public/private split defines exactly what outside auditors can rerun without crossing the institutional boundary or rerunning provider calls.

Two public checks keep richer semantics separate from the official estimate. First, a richer-policy activation check evaluates a candidate surface over the released corpus. It scans \ValCorpusIncidentCount{} incidents, defines 14 candidate actions and 18 candidate rules, and exercises all supported repair modes: \texttt{insert}, \texttt{insert\_before}, \texttt{remove}, and \texttt{defer\_to\_human\_approval}. The ordering probe identifies three deterministic repairs in a single chain. This activation evidence, summarized in Table~\ref{tab:policy_coverage}, shows that richer mandatory and ordering behavior is checkable and not pooled into the official provider estimate. Second, the public-metric approval sensitivity check treats the \ValOfficialRemovedActionCount{} removed approval-gated actions as deferred rather than suppressed. Deferral preserves task coverage across all models/arm cells but lowers precision and Jaccard scores relative to removal, as expected when proposed actions are retained for analyst approval.

\begin{table}[htpb]
\caption{\label{tab:policy_coverage}Policy coverage separates the reported provider estimate from richer-rule activation checks.}
\centering
\footnotesize
\renewcommand{\arraystretch}{1.08}
\setlength{\tabcolsep}{2pt}
\rowcolors{2}{stagefill}{white}
\begin{tabularx}{\columnwidth}{@{}>{\raggedright\arraybackslash}p{0.25\columnwidth}>{\centering\arraybackslash}p{0.10\columnwidth}>{\centering\arraybackslash}p{0.10\columnwidth}>{\raggedright\arraybackslash}p{0.25\columnwidth}>{\raggedright\arraybackslash}X@{}}
\toprule
Surface & Actions & Rules & Activated & Role \\
\midrule
Official evaluation & 5 & 4 & R3, R4 & provider estimate \\
Richer-rule check & 14 & 18 & 17 candidate rules & activation check \\
\bottomrule
\end{tabularx}
\end{table}

The verifier is also a systems component in a narrower sense than generic agent-safety layers. The monitor-based enforcement tradition establishes that policies over event sequences can be enforced by automata that suppress or edit executions~\cite{schneider2000enforceable,ligatti2005editautomata}; Progent and AgentSpec instantiate this idea for LLM agents by intercepting tool calls and evaluating declarative constraints before execution~\cite{shi2025progent,wang2025agentspec}. SOCpilot shares the architectural instinct of enforcement outside the generative core, but applies it to a different governed object: the reviewed action plan submitted to an analyst for decision support. The plan-level action sequence is the surface shared across providers, paired against a human baseline, and releasable for public audit. That surface is why the action boundary, rather than the tool-call boundary, is the stable measurement point under the declared protocol.

The paired human baseline is a compliance anchor. It ties the same incidents to the same mapping contract and catalog, so omission, overlap, and disagreement can be read as workflow-level compliance phenomena. The substantive comparison is of proposal paths evaluated against a paired analyst-authored baseline under a single verifier.

The same method admits richer declared evaluation objects that activate additional rule families, expand the approved action space, and preserve a public evidence path that outside auditors can rerun without private access. Examples include mandatory insertion, ordering repair, approval deferral, and multi-step dependencies, such as forensic collection before restoration and before reconnection. The method also supports cross-institution and cross-sector case studies: another SOC can replace the canonical cases, declare its own rule subset, and reuse the verifier and the paired baseline. The reporting rule follows from the method: changed inputs, policy surface, or verifier behavior define a new evaluation object rather than a silent extension of the current estimate. In the language of applied replicability, later studies should widen the bounds of the same hypothesis rather than quietly replace it with a new one~\cite{olszewski2025replicability}.

For this reason, the current paper reports a single, well-scoped result rather than a generic benchmark leaderboard. It shows that, under the released protocol, plan-level enforcement is the stable control surface even when prompt-level guidance is not. Scenario diversity strengthens the research program when introduced as an explicitly richer evaluation object, rather than as a silent change pooled into a single estimate.

The reported evaluation changes how plan-level compliance should be stated. Prompt-level policy text and action-level policy compliance are different properties: under one declared prompt lineage, the same inline policy block changed Claude’s run-level violation rate from \ValOfficialClaudeZeroViolationRate{} to \ValOfficialClaudePolicyViolationRate{}, whereas it moved GPT only from \ValOfficialGptZeroViolationRate{} to \ValOfficialGptPolicyViolationRate{}. Reporting policy-compliance numbers without separating these layers risks attributing behavior to ``the LLM’’ that depends on a specific provider, prompt, and policy-text version.

The evaluation also shows why the declared object matters. The deterministic verifier removed \ValOfficialRemovedActionCount{} approval-gated actions across the official run without any run losing task coverage relative to its paired analyst-authored baseline. Across \ValStabilityRepeatCount{} reruns, the same fixed corpus, catalog, and rule file produced rates within \ValStabilityIncidentViolationRateMin{}–\ValStabilityIncidentViolationRateMax{}, even though the providers do not expose seed control. What is stable is what the protocol fixes. Releasing the catalog, policy file, mapping rules, run manifest, and paired-baseline bundle enables a reader to distinguish a stronger system from a weaker policy surface in future comparisons.

\section{Threats to Validity}
\label{sec:validity}

Task-to-action mapping is the primary construct validity surface, as mapping errors can bias downstream overlap and burden estimates. The paired human baseline is derived from analyst-authored SOAR workflow records, not newly elicited study responses. In the released audit, coverage is complete: there are \ValCorpusUnmatchedTaskCount\ unmatched tasks and \ValCorpusAmbiguousMatchCount\ ambiguous ties across \ValCorpusSensitivityTaskCount\ mapped tasks. The remaining mapping evidence is lexical rather than tie-based: \ValCorpusSingleKeywordMatchCount/\ValCorpusSensitivityTaskCount\ unique matches rely on single-keyword support. This dependence is localized: it concentrates in \texttt{collect\_forensics}, while \texttt{isolate\_host} relies entirely on multi-keyword support in the released manifests. The paired human baseline is therefore a workflow-level compliance reference grounded in recorded analyst work. The artifact keeps the mapping surface inspectable through per-incident mapping-support manifests, a stratified mapping-inspection packet, and the global assessment bundle, without redistributing private task text or requiring new humans to construct the baseline.

Utility metrics are most informative jointly. A zero task-coverage drop means deterministic enforcement did not remove baseline-covered steps beyond those the raw LLM had already omitted. Table~\ref{tab:utility_precision} shows that enforcement improves precision in all model/arm cells. The remaining gap is directional rather than diffuse: extra recovery and egress-block actions dominate disagreement with the paired human baseline. Zero out-of-catalog rate establishes syntactic contract compliance, and the policy burden measured here is catalog-valid but governance-sensitive action choice. The alternative \texttt{defer\_to\_human\_approval} mode remains implemented, whereas the official study reports \texttt{remove} as the primary mode because it yields the cleanest repair trace and the least workflow ambiguity in the public artifact. The public sensitivity report shows the expected trade-off: deferral preserves coverage while also allowing more LLM-only actions for analyst review. The official provider runs exercise removal, and the artifact separately exercises insertion, ordering, and deferral through public checks.

Approval evidence is interpreted at the public evidence boundary. The released package shows that an action is approval-gated under the declared catalog and policy set; it does not expose institution-internal approval logs or latent supervisor context. Some historically executed actions may therefore have been acceptable in a private context that the public artifact cannot reveal. The reported evaluation chooses the externally auditable reading by design: without explicit approval context in the public evidence surface, the verifier treats the action as non-compliant. This preserves auditability and prevents the public claim from inheriting private assumptions that outside readers cannot inspect.

A second construct validity surface is the abstraction loss between institution-internal playbooks and the released canonical packages. Prior work on incident-response playbooks shows that real response procedures are shaped by organizational actors, negotiation, and local context beyond what a single incident record exposes~\cite{schlette2024playbooks,woods2023lessonslost}. The public artifact omits that private context so that the released evidence surface remains auditable. The paper measures compliance under the exported plan/evidence surface: exactly the object that reviewers can inspect and future systems can reuse.

A third validity surface is protocol-surface compression. The declared action catalog contains five actions, and the empirically exercised provider slice is approval treatment over \texttt{restore\_host} and \texttt{isolate\_host}. This is the declared measurement object and the source of the paper’s official estimates. Richer mandatory chains, denser ordering dependencies, and a broader recovery vocabulary belong in the next declared evaluation object, where lower burden can be attributed only after the corpus, catalog, and rule surface are fixed.

\begin{table}[htpb]
\caption{\label{tab:utility_precision}Utility decomposition by model and arm. Precision rises after deterministic enforcement in all four cells.}
\centering
\footnotesize
\renewcommand{\arraystretch}{1.08}
\setlength{\tabcolsep}{3pt}
\rowcolors{2}{stagefill}{white}
\begin{tabularx}{\columnwidth}{@{}LCCC@{}}
\toprule
Model / arm & \shortstack[c]{Raw\\precision} & \shortstack[c]{Enforced\\precision} & $\Delta$Jaccard \\
\midrule
Claude / zero & \ValOfficialClaudeZeroPrecisionRaw & \ValOfficialClaudeZeroPrecisionEnforced & \ValOfficialClaudeZeroDeltaJaccard \\
Claude / policy & \ValOfficialClaudePolicyPrecisionRaw & \ValOfficialClaudePolicyPrecisionEnforced & \ValOfficialClaudePolicyDeltaJaccard \\
GPT / zero & \ValOfficialGptZeroPrecisionRaw & \ValOfficialGptZeroPrecisionEnforced & \ValOfficialGptZeroDeltaJaccard \\
GPT / policy & \ValOfficialGptPolicyPrecisionRaw & \ValOfficialGptPolicyPrecisionEnforced & \ValOfficialGptPolicyDeltaJaccard \\
\bottomrule
\end{tabularx}
\end{table}

The \texttt{asset\_criticality} is currently a proxy for incident severity copied during canonicalization, so severity- and criticality-stratified summaries are contextual rather than explanatory. The verifier measures adherence to the reported policy set. Inconsistent rule sets (self-dependencies and ordering cycles) fail fast and are escalated rather than silently repaired. Under fixed rules and a fixed action catalog, the verifier acts as a deterministic finite repair transducer over action sequences. Our claim is therefore bounded and traceable: policy treatment over one declared surface, not a general-purpose policy reasoner.

External validity is anchored to the provider state and institutional setting. In the official evaluation, the two providers respond in opposite directions to policy prompting. Table~\ref{tab:evaluation_pairwise_tests} shows the within-model Claude contrast surviving Holm correction with large marginal separation (Cohen’s $h = 1.12$), whereas the within-model GPT contrast also survives correction but at $h = -0.14$, an effect size small enough to be read as borderline-detectable rather than substantively important. The result is asymmetric and useful: policy text worsens Claude in the reported evaluation, while its effect on GPT is too small to be operationally meaningful. The lesson is architectural rather than provider-specific: policy-aware text is not the same as policy-conformant action selection.

The design measures provider divergence at the action boundary rather than attributing it to hidden mechanisms in the model. Provider-side guardrails, instruction hierarchy handling, refusal style, and latent planning preferences may all contribute. The contribution is behavioral and operational: under a single declared prompt lineage and a shared verifier path, policy text alone did not yield stable action-level compliance across providers.

Rule frequencies and disagreement patterns depend jointly on the incident mix, the exposed action surface, and the approval evidence available in the released package. A corpus with more restoration-heavy cases, richer host-level approval traces, or broader mandatory-rule activation would define a different estimate. The present result is an auditable slice with declared boundaries, which is precisely what enables comparison and extension.

Comparison must therefore be declared rather than silent: any change to the incident slice, mapping contract, policy file, prompt template, provider endpoint, or model identifier creates a separate evaluation object rather than a direct extension of the current estimates.

\section{Related Work}
\label{sec:related-work}

Prior work addresses related aspects of this problem across four research lines, each of which stops at a boundary that SOCpilot crosses: workflow characterization, event-level detection, model capability assessment, and runtime agent control. No adjacent work combines a real SOC corpus, typed policy treatment, a paired human baseline, and a fixed public verification surface.

\noindent\textit{SOC operations and governance.}
Interviews and field-study work examine SOC organizational factors. Kokulu et al.\ study how institutional alignment between security and business units shapes incident-response processes through qualitative interviews across SOCs of varying maturity~\cite{kokulu2019matched}, whereas Jones et al.\ document the operational pressures on endpoint management introduced by the COVID-19 work-from-home (WFH) transition~\cite{jones2023wfh}. At the alert layer, Vermeer et al.\ (AsiaCCS 2022) track network intrusion detection system (NIDS) ruleset evolution longitudinally, showing that a small fraction of rules accounts for the majority of alerts~\cite{vermeer2022rulingrules}, and Yang et al.\ measure the distribution of true attacks, attack attempts, and benign triggers in four years of production traffic~\cite{yang2024trueattacks}. Vermeer et al.\ (CCS 2023) characterize the decision workflows by which SOC analysts triage alerts into actionable incidents~\cite{vermeer2023alertalchemy}. Woods et al.\ further demonstrate that legal, insurance, and coordination pressures imposed by external stakeholders constrain which incident-response actions an organization can safely take~\cite{woods2023lessonslost}. Closest in spirit to SOCpilot, Schlette et al.\ measure the gap between documented playbook steps and the actions analysts actually select across a multi-SOC study~\cite{schlette2024playbooks}. However, none of these works provides a mechanism to enforce or verify compliance, nor do they produce a quantitative measurement of plan-level policy adherence against a paired human baseline under a fixed public evidence surface.

\noindent\textit{Provenance, detection, and reconstruction.}
Event-level reasoning work addresses a different layer. Altinisik et al.\ and Zhang et al.\ apply graph representation learning and task-guided segmentation, respectively, to search and detect advanced persistent threat (APT) behavior within system-audit-log provenance graphs~\cite{altinisik2023provg,zhang2025tapas}. Dong et al.\ and Aly et al.\ reconstruct attack stories from endpoint detection records~\cite{dong2023pedr,aly2025ocrapt}. These works answer \emph{what happened} in an incident, not whether the subsequent response plan satisfies organizational policy constraints.

\noindent\textit{LLM assistance in cybersecurity tasks.}
Deng et al.\ build a penetration-testing agent and evaluate it on structured exploitation tasks~\cite{deng2024pentestgpt}; Stafeev et al.\ extend the goal-directed agent paradigm to task-driven web-application scanning~\cite{stafeev2025yurascanner}; Deng et al.\ (RACONTEUR) evaluate LLMs on security-report generation~\cite{deng2025raconteur}; Wang et al.\ systematically benchmark LLM security knowledge and multi-step reasoning against human analyst baselines on a domain-specific question set~\cite{wang2025digitalexpert}. CyberSecEval~\cite{bhatt2024cyberseceval} provides suite-level capability assessment at the prompt and code levels. All of these works measure what a model can produce in an open-task setting; none measure whether the proposed action sequence satisfies a typed, institution-specific policy when applied to real incidents using a fixed action catalog.

\noindent\textit{LLM serving-layer risk and runtime control.}
Measurement work on prompt leakage, prompt inversion, plugin trojaning, and cross-app isolation identifies attack surfaces in LLM-integrated deployments~\cite{wu2025promptleakage,qu2025promptinversion,dong2025philosophers,wu2025isolategpt}. On the defensive side, Shi et al.\ (Progent) introduce a programmable privilege-control layer that intercepts tool calls before execution and enforces security policies specified in a domain-specific language~\cite{shi2025progent}, and Wang et al.\ (AgentSpec) provide a declarative constraint language for bounding autonomous agent behavior across planning steps~\cite{wang2025agentspec}. Recent and concurrent systems broaden this design space: ToolGuard-style guard generation compiles business-policy documents into tool-level checks~\cite{zwerdling2025toolguard}, Policy Compiler for Secure Agentic Systems (PCAS) compiles authorization policies into an instrumented agentic system with a reference monitor over dependency graphs~\cite{palumbo2026pcas}, VeriGuard synthesizes and formally verifies policy code before runtime monitoring~\cite{miculicich2025veriguard}, ShieldAgent reasons over action trajectories with verifiable policy circuits~\cite{chen2025shieldagent}, and LogiSafetyGen evaluates implicit regulatory compliance in tool-invocation traces~\cite{song2026logisafetygen}. These systems reinforce the same architectural lesson: policy treatment should sit outside model text alone, and they govern live agents, tool calls, generated guard code, or synthetic tool-use traces. SOCpilot’s verifier governs a different object: the reviewed incident-response \emph{plan}, submitted by an assistant that does not execute tools directly. The action sequence is the only surface shared across providers, paired human baseline, and public artifact, so enforcement operates pre-execution at the plan boundary rather than at the tool-call boundary during live execution.

\noindent\textit{Monitor-based enforcement.}
Schneider and Ligatti et al.\ establish the theoretical foundation for safety properties enforceable by automata-based monitors that suppress or edit executions~\cite{schneider2000enforceable,ligatti2005editautomata}. Kim et al.\ (Kinetic) and Yu et al.\ (Precise Security Instrumentation, PSI) instantiate these ideas in operational software-defined networking (SDN) and network-policy systems~\cite{kim2015kinetic,yu2017psi}. Collaborative Automated Course of Action Operations (CACAO) standardizes cybersecurity playbooks as structured workflow objects with typed branching and parallelism~\cite{jordan2023cacao}. SOCpilot applies the monitor intuition to a new governed object: pre-execution incident-response plans over a declared action catalog. The same typed verifier path is applied to both LLM proposals and paired human baselines, with all results anchored to a fixed, publicly verifiable evidence surface.

\section{Conclusion}
\label{sec:conclusion}

SOCpilot establishes that policy compliance in LLM-assisted incident-response planning is a property of the proposed action trace, not of the prompt that generated it. Across a fixed evaluation object, identical policy text induces divergent provider behavior, showing that policy-aware generation does not imply policy-compliant action selection. By moving enforcement to a deterministic layer over typed action sequences, SOCpilot turns compliance into a verifiable property at the plan-review boundary, independent of model internals, prompt formulation, or provider-specific behavior. This reframes how compliance claims should be made and evaluated: not as attributes of model outputs in isolation, but as properties of a declared, auditable object that binds incident context, action vocabulary, policy semantics, and enforcement. For LLM-assisted security operations, the relevant contribution is whether the plan an analyst is about to approve satisfies the governing rules under a fixed, inspectable protocol.

\section*{Ethical Considerations}
\label{sec:ethical-considerations}

This study uses real incident records from an anonymized production SOC in a financial-sector case study. The primary ethical risks are privacy leakage, unintended disclosure of operational procedures, and overclaiming beyond evidence from a single institution and a single declared policy/action space. These risks are mitigated by enforcing a strict private/public boundary: raw exports remain institution-internal, anonymization precedes any public release, and the released artifact contains only canonical packages and non-private analysis outputs. Anonymization followed a dual-review process, with operational review before export and an independent research-side rerun with PII scanning, forbidden-term checks, and manual verification. The study evaluates only pre-execution plan compliance: the LLM does not execute tools; the verifier acts before any response action is taken; and the paper limits its strongest claims to the current approval-gated rule slice rather than to universal cross-institutional safety.

\bibliographystyle{IEEEtran}
\bibliography{references}

\appendix

\section*{Open Science}
\label{sec:open-science}
\looseness=-1

The paper’s main claims are carried in the body. This appendix is supplementary: it states the public audit question, records the interpretation rules for the declared evaluation state, and preserves secondary quantitative views that would interrupt the main argument.

The artifact supports one external audit question: Can an independent evaluator confirm that the released inputs, manifests, public run accounting, aggregate outputs, and public rechecks substantiate the reported claims under the declared evaluation state? This is narrower than a full institutional replay, but it is the right question for a release that excludes raw case lineage, private approval context, and provider reruns.

Within that boundary, the bundle exposes the action vocabulary, typed policy file, mapping-support evidence, official analysis summaries, public run accounting, paired contrasts, and human-baseline analysis needed to rederive the public results. The release prioritizes transparent re-analysis of the declared study before any attempt to widen the hypothesis to new providers, policies, or incident slices~\cite{olszewski2025replicability}.

The same discipline matters for future benchmark design. Provider names, prompt arms, the declared action catalog, the active rule slice, the paired-baseline contract, and the repair modes allowed by the verifier must be fixed before a lower violation rate can be read as a comparable improvement. The paired human baseline is included for the same reason: a verifier-only release could show correction under one rule set, but it could not show whether surviving plans align with recorded workflow patterns from the same incidents.

\subsection*{Claim Surface and Verification Workflow}

The public bundle anchors three claim families. Corpus release and boundary claims are supported by canonical incident packages and declared-input manifests. Violation, enforcement, utility, execution accounting, and paired contrasts are supported by the official metrics, aggregate summaries, public run accounting, and paired-test outputs. Baseline validity and surface fragility are supported by mapping-support manifests and the global-artifact assessment. Provider calls, raw provider payloads, host-level ground truth, and behavior under new prompts, models, or incident slices are out of scope. Detailed file inventories, schemas, and reproduction commands belong in the artifact README, not in the paper.

Five interpretation rules govern claims made against the declared evaluation state. A new provider result under the same incident slice, mapping contract, policy set, action catalog, and public audit surface is directly comparable within the current evaluation object. A change to any of those elements creates a separate evaluation object rather than a silent improvement over the current one. The paired human baseline is a second auditable compliance reference under the same slice and mapping contract; it is not a universal optimum or an abstract provider-versus-human ranking. The verifier is the last pre-execution control point over policy treatment, not a substitute for analyst reasoning or a general planner. Finally, public validation of the shipped non-private analysis bundle constitutes the external audit of the paper's strongest public claim, whereas host-level outcome truth, institution-internal lineage, and provider reruns lie outside the current public evidence surface.

Public reproduction is documented in the artifact README. The release workflow confirms the packaged files and their hashes, regenerates public assessment outputs, regenerates the paper result assets, and writes a reproduction report. It does not rerun LLM calls, but it reproduces the public analyses and audits the evidence surface on which the main claims rest.

\section*{Use of Generative Artificial Intelligence Tools}
\label{app:ai-tools}

The authors used Grammarly and ChatGPT for grammar checking, limited editorial revision, and minor formatting assistance for figures and illustrations. All scientific claims, analyses, experiments, visual representations, and conclusions were produced and verified by the authors.

\subsection*{Supplementary Quantitative Detail}
The appendix keeps only quantitative views that support audit decisions, but would interrupt the main argument in the body. Table~\ref{tab:execution_diagnostics} preserves the per-cell repair burden and execution accounting behind the rule-concentration discussion. Table~\ref{tab:approval_mode_sensitivity} preserves the public sensitivity view for approval deferral. Table~\ref{tab:pilot_calibration} preserves the pooled across-provider screening view; the body focuses on provider-resolved rates because the two providers move in opposite directions under policy prompting. Table~\ref{tab:evaluation_rule_treatment} preserves the exact rule-level treatment rates behind the approval-slice discussion. These appendix-only objects extend numeric traceability without changing the interpretation established in the main text.

\begin{table}[htpb]
\caption{\label{tab:execution_diagnostics}Execution and rule-burden diagnostics by provider cell.}
\centering
\footnotesize
\renewcommand{\arraystretch}{1.08}
\setlength{\tabcolsep}{2pt}
\rowcolors{2}{stagefill}{white}
\begin{tabularx}{\columnwidth}{@{}>{\raggedright\arraybackslash}p{0.24\columnwidth}CCCC>{\centering\arraybackslash}p{0.13\columnwidth}@{}}
\toprule
Model / arm & Complete & R3 & R4 & \shortstack[c]{Out-of-\\catalog} & Failures \\
\midrule
Claude / policy & 200/200 & 173 & 9 & 0 & 0 \\
Claude / zero & 200/200 & 65 & 8 & 0 & 0 \\
GPT / policy & 200/200 & 89 & 9 & 0 & 0 \\
GPT / zero & 200/200 & 104 & 9 & 0 & 0 \\
\bottomrule
\end{tabularx}
\end{table}

\begin{table}[htpb]
\caption{\label{tab:approval_mode_sensitivity}Approval deferral sensitivity from public metrics.}
\centering
\footnotesize
\renewcommand{\arraystretch}{1.08}
\setlength{\tabcolsep}{2pt}
\rowcolors{2}{stagefill}{white}
\begin{tabularx}{\columnwidth}{@{}>{\raggedright\arraybackslash}p{0.22\columnwidth}>{\centering\arraybackslash}p{0.15\columnwidth}>{\centering\arraybackslash}p{0.19\columnwidth}>{\centering\arraybackslash}p{0.23\columnwidth}>{\centering\arraybackslash}p{0.14\columnwidth}@{}}
\toprule
Model / arm & Deferred & \shortstack[c]{Precision\\remove/defer} & \shortstack[c]{Jaccard\\remove/defer} & \shortstack[c]{Coverage\\loss} \\
\midrule
Claude / policy & 182 & 0.927/0.667 & 0.705/0.550 & none \\
Claude / zero & 73 & 0.945/0.827 & 0.725/0.657 & none \\
GPT / policy & 98 & 0.973/0.811 & 0.722/0.630 & none \\
GPT / zero & 113 & 0.979/0.807 & 0.791/0.682 & none \\
\bottomrule
\end{tabularx}
\end{table}

\begin{table}[htpb]
\caption{Pooled official evaluation by arm. Provider-resolved rates appear in Figure~\ref{fig:gate2_violation_rates}.}
\label{tab:pilot_calibration}
\centering
\footnotesize
\renewcommand{\arraystretch}{1.08}
\setlength{\tabcolsep}{3pt}
\rowcolors{2}{stagefill}{white}
\begin{tabularx}{\columnwidth}{@{}>{\raggedright\arraybackslash}p{0.21\columnwidth}>{\centering\arraybackslash}p{0.08\columnwidth}>{\centering\arraybackslash}p{0.13\columnwidth}>{\centering\arraybackslash}p{0.08\columnwidth}>{\centering\arraybackslash}p{0.20\columnwidth}>{\centering\arraybackslash}p{0.17\columnwidth}@{}}
\toprule
Arm & Runs & \shortstack[c]{Hard\\count} & \shortstack[c]{Violation\\rate} & \shortstack[c]{95\%\\interval} & \shortstack[c]{Verifier\\edit rate} \\
\midrule
LLM zero &
\ValPilotLlmZeroRunCount &
\ValPilotLlmZeroHardViolationCount &
\ValPilotLlmZeroViolationRate &
\ValPilotLlmZeroViolationCiLow--\ValPilotLlmZeroViolationCiHigh &
\ValPilotLlmZeroEnforcementModRate \\
LLM policy &
\ValPilotLlmPolicyRunCount &
\ValPilotLlmPolicyHardViolationCount &
\ValPilotLlmPolicyViolationRate &
\ValPilotLlmPolicyViolationCiLow--\ValPilotLlmPolicyViolationCiHigh &
\ValPilotLlmPolicyEnforcementModRate \\
\rowcolor{stagefill}
Human baseline &
\ValPilotHumanRunCount &
\ValPilotHumanHardViolationCount &
0 &
0--\ValPilotHumanViolationCiHigh &
0 \\
\bottomrule
\end{tabularx}
\end{table}

\begin{table}[htpb]
\caption{Official evaluation rule-level treatment comparison (zero vs.\ policy prompt). Bold marks the dominant rule-family effect.}
\label{tab:evaluation_rule_treatment}
\centering
\footnotesize
\renewcommand{\arraystretch}{1.08}
\setlength{\tabcolsep}{2pt}
\rowcolors{2}{stagefill}{white}
\begin{tabularx}{\columnwidth}{@{}LCCCC@{}}
\toprule
Model/rule & Zero & Policy & $\Delta$ & Runs \\
\midrule
\textbf{claude / R3} & \textbf{0.3250} & \textbf{0.8650} & \textbf{+0.5400} & \textbf{200} \\
claude / R4 & 0.0400 & 0.0450 & +0.0050 & 200 \\
GPT / R3 & 0.5200 & 0.4450 & $-$0.0750 & 200 \\
GPT / R4 & 0.0450 & 0.0450 & 0 & 200 \\
\bottomrule
\end{tabularx}
\end{table}

\end{document}